\documentclass[twocolumn]{aastex631}

\usepackage{enumerate}
\usepackage{subfigure}
\usepackage{float}
\usepackage{natbib}
\setcitestyle{square, comma, numbers,sort&compress, super}
\usepackage{amsmath}
\pdfoutput=1

\begin{document}

%\title{Lithium enhancement absent in secondary red clump stars}
%\title{Survey for lithium-rich giants among low- and high mass red clump giants:  An evidence for He-flash as the cause for Li enhancement among low mass Li-rich red clump giants  }
\title{ Lithium abundances in giants as a function of stellar mass:  An evidence for He-flash as the source of Li enhancement in low mass giants  }

\author[0000-0002-4282-605X]{Anohita Mallick}
\affiliation{Indian Institute of Astrophysics, 560034, 100ft road Koramangala, Bangalore, India}
\affiliation{Pondicherry University, R. V. Nagara, Kala Pet, 605014, Puducherry, India}

\author[0000-0001-8360-9281]{Raghubar Singh}
\affiliation{Aryabhatta Research Institute of Observational Sciences (ARIES), Manora Peak, Nainital 263001, India}

\author[0000-0001-9246-9743]{Bacham E. Reddy}
\affiliation{Indian Institute of Astrophysics, 560034, 100ft road Koramangala, Bangalore, India}



\begin{abstract}
In this work, we studied the distribution of lithium abundances in giants  as a function of stellar mass. We used a sample of 1240 giants common among Kepler photometric and LAMOST medium resolution (R $\approx$ 7500) spectroscopic survey fields. The asteroseismic $\Delta$P - $\Delta \nu$ diagram is used to define core He-burning red clump giants and red giant branch stars with inert He-core. Li abundances have been derived using spectral synthesis for the entire sample stars. Directly measured values of asteroseismic parameters $\Delta$P(or $\Delta \Pi_1$) and $\Delta \nu$ are either taken from the literature or measured in this study. Of the 777 identified red clump giants, we found 668 low mass ($\leq$ 2~M$_{\odot}$) primary red clump giants and 109 high mass ($>$ 2~M$_{\odot}$) secondary red clump giants. Observed Li abundances in secondary red clump giants agree with the theoretical model predictions. The lack of Li-rich giants among secondary red clump giants and the presence of Li-rich, including super Li-rich giants, among primary red clump stars reinforces the idea that Helium-flash holds the key for Li enrichment among low-mass giants. The results will further constrain theoretical models searching for a physical mechanism for Li enhancement among low-mass red clump giants. Results also serve as observational evidence that only giants with mass less than $\approx$ 2~M$_{\odot}$ develop degenerate He-core and undergo He-flash. 
 
\end{abstract}


\section{Introduction} \label{sec:intro}
It is well understood that stars on the red giant branch (RGB) undergo convective mixing or the first dredge-up, which alters photospheric abundances of several elements like helium, carbon, nitrogen, and lithium.  
Of those, Li is the most affected element. The surface Li abundance, A(Li), drops to $\sim$ 95\% of their main-sequence value \citep{1967Iben}. The discovery of the first Li-rich giant (LRG) \citep{1982Wallerstein} four decades ago challenged the general understanding of Li evolution. Since then, numerous works \citep{1989Brown, 2011Kumar, 2013Martell, 2019Deepak, 2019Casey} established that a small fraction ($\sim$~1~\%) of giants exhibit high levels of Li abundances compared to standard model predictions of A(Li) = 1.5 - 1.8~dex depending on mass. In many cases, the enhancement surpasses the primordial A(Li) $\simeq$ 2.7 dex, and in a few cases, A(Li) is more than the present interstellar medium value of $\sim$ 3.2~dex which is tagged as super lithium-rich (SLR) stars. The existence of such stars transcends the standard evolutionary theories implying some ancillary mechanism behind the production and preservation of Li in LRGs. Various theories were proposed to explain the mechanism and site of Li production in giants which include  in-situ production and external origin \citep{1995Boothroyd, 2000Charbonnel, 2007Charbonnel, 2009Denissenkov, 2012Carlberg}. 

\begin{figure*}
\subfigure[]{\label{fig:1a}\includegraphics[width=\columnwidth,height=6.9cm]{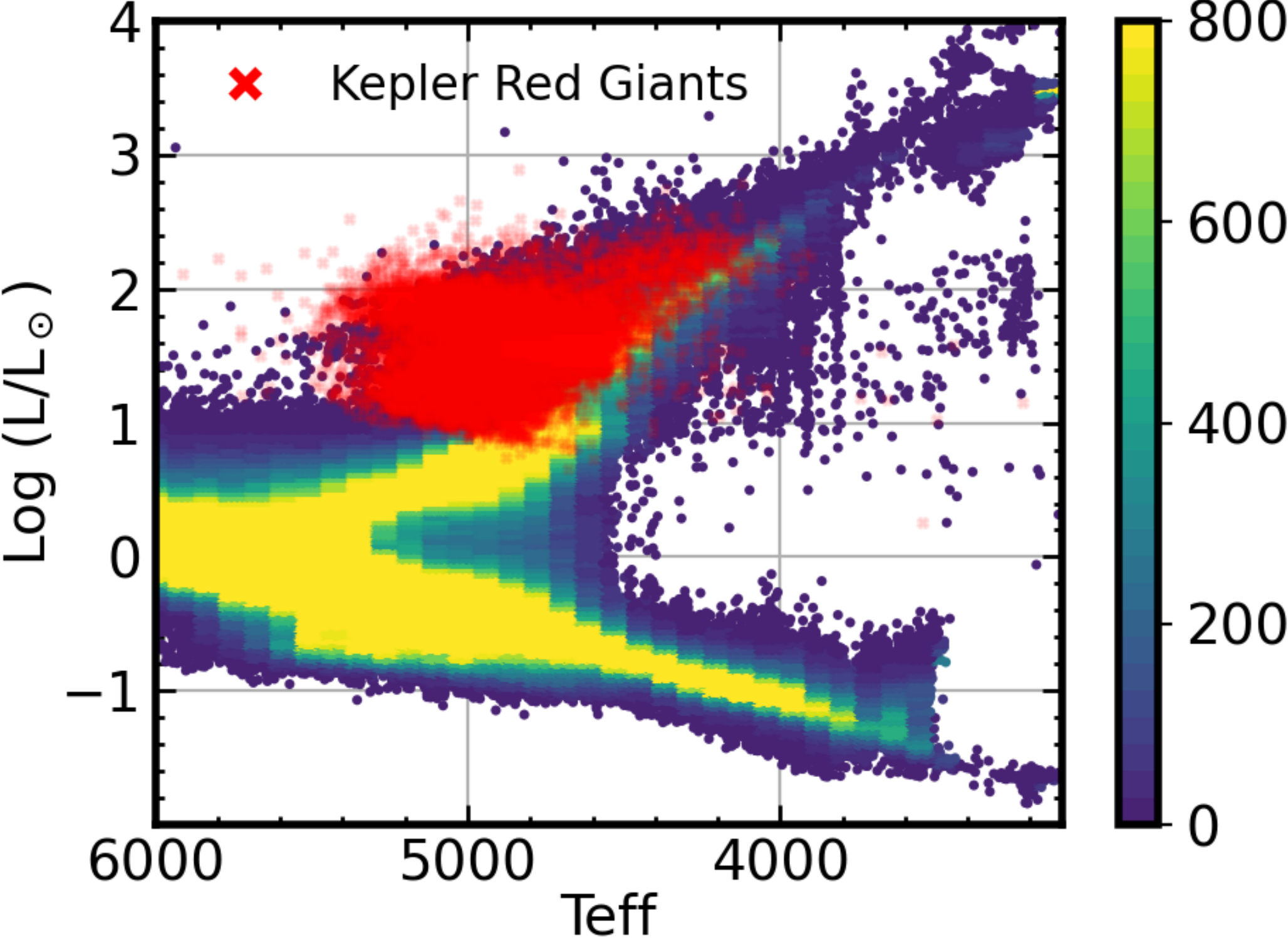}}
\subfigure[]{\label{fig:1b}\includegraphics[width=\columnwidth]{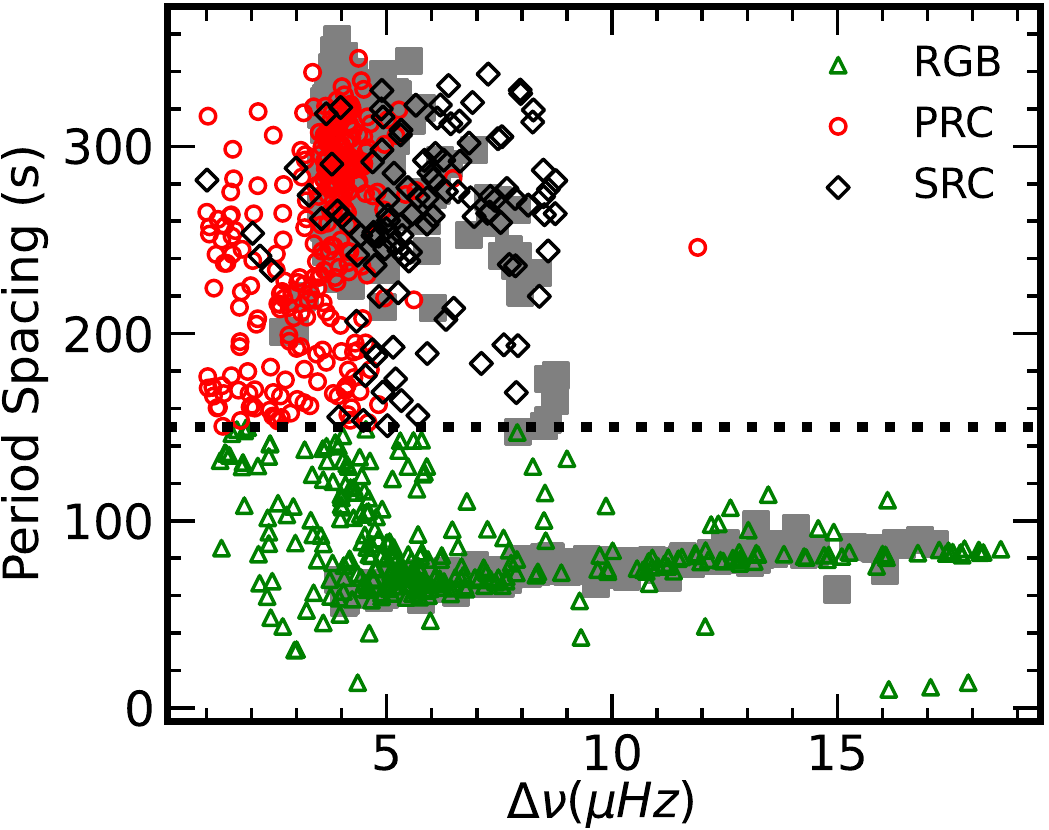}}
\caption{ Shown are the sample giants of RC and RGB. In panel (a), 16,094 stars (red symbols) from \citet{2018Yu} are plotted with the entire Kepler sample as background. In panel (b), asteroseismically classified RGB and RC sample stars in this study are shown along with the sample giants from \citet{2016Vrard}.}
\end{figure*}

\par
While the Li origin debate was underway, a large observational study by \citep{2011Kumar}  hypothesized that Li enhancement might be associated with the red clump (RC) stars with a helium-burning core. Lithium synthesized in the interior by a process known as Cameron-Fowler mechanism \citep{1971Cameron} might have mixed up with the outer layers by non-canonical processes during the Helium flash - a runaway nuclear burning  at the RGB tip in low-mass stars ($\leq$ 2~M$_{\odot}$). Now hundreds of LRGs exist in the literature primarily due to large systematic studies \citep{2019Deepak, 2019Singh, 2019Casey,2021Yan,2021Martell} based on large spectroscopic surveys such as LAMOST and GALAH and the space-based data sets of Gaia astrometry and time-resolved photometry from  Kepler and TESS. Also, these studies concluded that most LRGs belong to the He-core burning RC phase.  The study by \citet{2020Kumar} demonstrated that Li enhancement is ubiquitous among core He-burning low-mass RC giants, and Li gets only depleted in giants ascending the RGB implying He-flash at the RGB tip is the most likely cause  for Li enhancement among RC giants. Further, \cite{2021Singh}, using Li abundances and asteroseismic parameters, showed that most of the SLR giants are very young RCs compared to Li-normal RC giants implying Li enhancement in SLRs occurred very recently. A recent observational study by \citet{2022Sneden} showed that the Li-rich giants are more likely to have a strong chromospheric He line at 10830~\AA\, opening a new avenue to probe Li-rich origin. These observational results are yet to be supported by  theoretical modeling. One of the physical mechanisms proposed for Li enhancement is extra mixing  due to the excitation of internal gravity waves by turbulent convection caused by He-flash and the associated large luminosity \citep{2020Schwab}. 
\par 
There are still questions related to high Li origin in giants. Is the He-flash a universal production mechanism for Li? Does Li enhancement occur only in Low mass giants, or are there massive ($>$ 2~M$_{\odot}$) Li-rich giants as well? Till now, surveys  focused mostly on low-mass RC giants at the cooler end of the horizontal branch (HB).  
It would be interesting to map LRGs over a range of masses to understand the origin of Li enhancement. This is because  if He-flash is the sole mechanism for high Li in giants, one wouldn't expect LRGs among high-mass giants as He-flash is expected to occur only in low mass ($\leq$ 2~M$_{\odot}$) giants. Recently, %there was an attempt by \cite{2021Deepak} to study Li distribution as a function of mass. However, they couldn't draw any conclusion as their sample is relatively much smaller. On the other hand,
the two observational  studies \citep{2021Martell, 2022Zhou} showed the presence of LRGs across mass and evolutionary phases implying Li enhancement in giants could be from multiple sites. However, both studies suffer from ambiguity in determining giants' evolutionary phase due to a lack of direct measurement of asteroseismic parameters. 

\par
Here, we assembled one of the largest data sets of 1240 giants for which direct measurements of asteroseismic parameters and LAMOST medium-resolution spectra are available. The maximum initial mass of a star (M$_{\rm HeF}$) to experience He-flash ranges from 1.8 - 2.2~M$_{\odot}$ depending on its metallicity \citep{1992Chiosi}. In this study, we probed Li abundance patterns as a function of mass among  RC giants. In particular, among the ``secondary RC stars" using a cut-off mass M$_{\rm HeF}$ $>$ 2~M$_{\odot}$. 

\section{Sample Selection} \label{sec:sample}
It is challenging to unambiguously identify giants' evolutionary phase, particularly the He-core burning  RC giants from those ascending the RGB solely by their location in the $T_{\rm eff} - L$ diagram (HRD). It is because the luminosity bump and upper RGB  overlap with the red clump regions. It is essential to have an independent way of knowing stars' evolutionary phases, either as RC or RGB giants. The Kepler space mission data provides high-quality and high-cadence time-resolved photometry required for asteroseismic analysis. RC and RGB giants show  characteristic oscillation properties enabling one to separate RC giants from those on RGB \citep{Bedding2011}. 

\par
We adopted a sample of 16,094 giants from the catalogue by \citet{2018Yu}, who compiled giants for which oscillations were detected.  It  contains only stars with log $g$ $\geq$ 1.5 or log ($L/L_{\odot}$) $\leq$ 2.24 dex i.e., only RC and RGB giants. The catalogue excluded stars with the frequency of maximum oscillation power, $\nu_{\rm max}$ $<$ 5 $\mu$Hz and $\nu_{\rm max}$ $\geq$ 275 $\mu$Hz. The lower value culls out super giants, and the higher limit excludes dwarfs and sub-giants. This sample is shown in the HRD (Fig. \ref{fig:1a}) with the entire Kepler data as background. The sample's luminosity values are estimated using stellar parallax, and values of apparent magnitude are taken from the Gaia DR3 catalogue. One could notice giants in the well-defined RC region in the HRD. \citet{2018Yu} provided asteroseismic parameters ($\nu_{\rm max}$, $\Delta\nu_{\rm max})$ and stellar parameters (mass, radius, log $g$, $[Fe/H]$) for the entire sample based on homogeneous asteroseismic analysis. 

Our main focus in this study is to understand  the distribution of Li among giants as a function of mass. For this, we cross-matched the \citet{2018Yu} sample with the LAMOST medium resolution spectroscopic survey DR7 catalogue \citep{2022Luo}. We found 1240 giants common between the two catalogues. All the giants have reasonably good spectra with Signal-to-Noise (SNR) $\geq$ 30. Most of the spectra (88~$\%$) have S/N $>$ 50. For the classification of giants as RC or RGB we turned to literature because \citet{2018Yu} catalogue did not provide mixed mode period spacing  ($\rm \Delta P$) values. Their classification is based on $\nu_{\rm max}$ and $\Delta \nu$ ( see their Fig~7). We have not adopted this method as there may be a possibility of  contamination of RCs with RGB or vice versa, particularly at lower $\Delta \nu$ ($<$ 10 $\mu$Hz). To minimize contamination, we adopted directly measured values of $\rm \Delta \Pi_1$ from the literature \citep{2016Vrard, 2019Singh}. We found $\Delta \Pi_1$ values for 584 giants. For the remaining 656 giants, the $\rm \Delta P$ values are calculated in this study using Kepler light curves (see appendix \ref{appendix:A}). The 1240 giants will be our working sample for this study, and the sample is shown in the asteroseismic diagram of $\rm \Delta P$ vs $\Delta \nu$ along with the sample of RC and RGBs classified  by \cite{2016Vrard} as background. Going by the convention, we classified giants with $\Delta P$(or $\Delta \Pi_1$) $\geq$ 150~s as  RC giants and giants with $\rm \Delta P$(or $\Delta \Pi_1$) $<$ 150~s as RGB giants. The $\rm \Delta P$ demarcation divides the sample into 777 RC giants and 463 RGB giants. Small contamination can't be ruled out as there is a small overlap of RC and RGB space in the $\rm \Delta P$ - $\Delta \nu$ diagram, particularly at ~$\Delta \nu \approx 5\ \mu Hz$. However, as the sample in Fig. \ref{fig:1b} suggests, the RC sample with $\rm \Delta P$ $\geq$ 150~s may be least contaminated with RGB sample as the RC sample cut-off at $\rm \Delta P$ is $\approx$ 3-$\sigma$ away from the mean trend of RGB in the plot. 
%We cannot say the same for RGB, particularly at the lower $\Delta \nu$ regime, RGB seems to be contaminated relatively more with the RC giants. 

\begin{figure*}
\includegraphics[width=\textwidth]{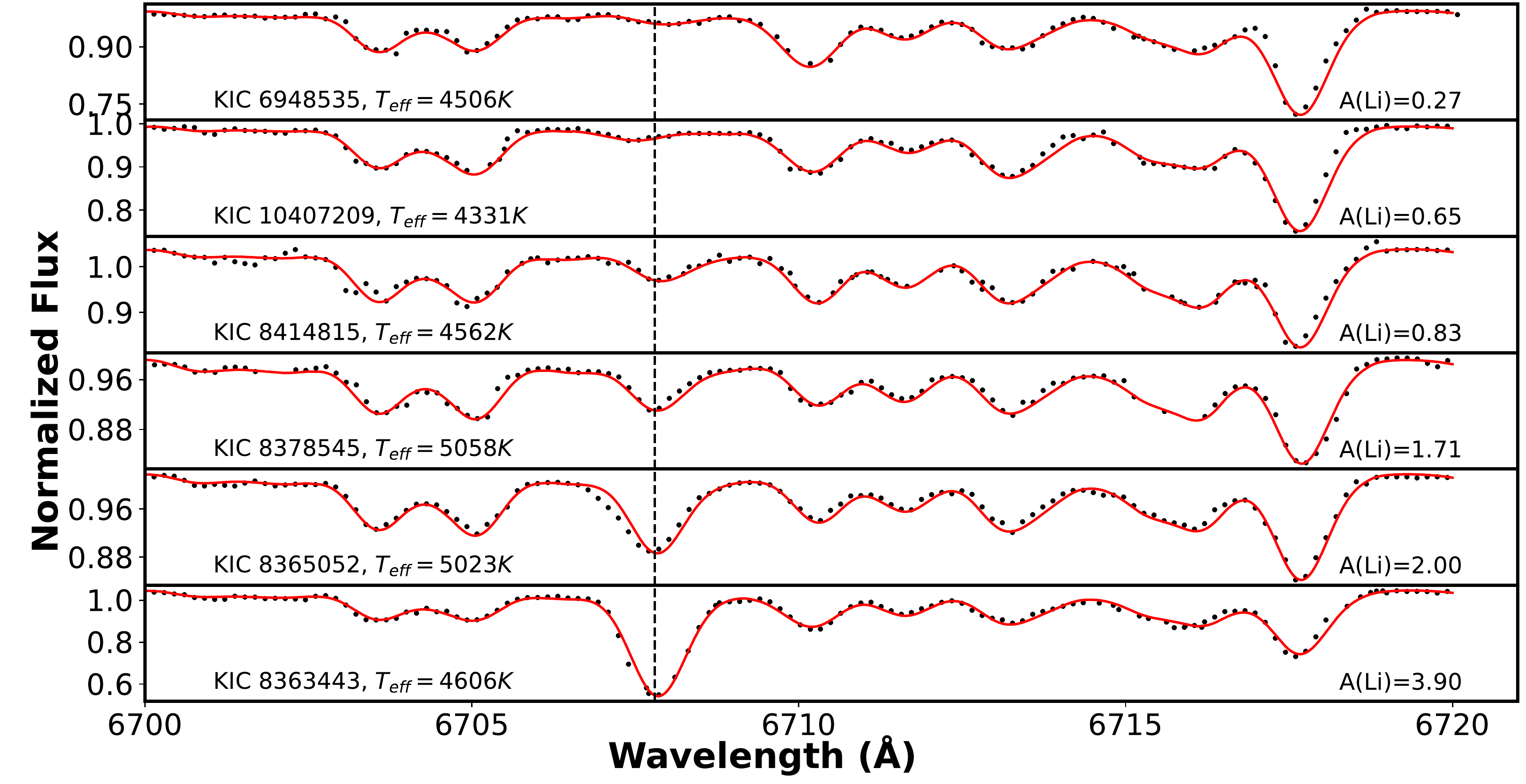}
\caption{Determination of Li abundances for a few representative giants using spectrum synthesis around Li 6707.8 \r{A} resonance line.
\label{fig:2}}
\end{figure*}

\section {Analysis}
Our primary goal in this study is to understand Li abundances among RC giants with a range of stellar masses. 

\subsection{Stellar mass Estimation}
The traditional method of using evolutionary tracks for mass determination may not yield desired results as tracks degenerate at RC and RGB regions. The scaling relations based on asteroseismic parameters are found to be useful for individual stellar masses. The study by \citet{2018Yu} provides estimated masses using the revised scaled relations \citep{2016Sharma}. Here, we provide a brief account:

\begin{equation}
    \frac{M}{M_{\odot}} =\left(\frac{\nu_{max}}{f_{\nu_{max}}\nu_{max,\odot}}\right)^3 \left(\frac{\Delta\nu}{f_{\Delta\nu}\Delta\nu_{\odot}}\right)^{-4}\left(\frac{T_{eff}}{T_{eff,\odot}}\right)^{1.5}
\end{equation}

here $f_{\nu_{\rm max}}$ and $f_{\Delta\nu}$ are correction factors for $\nu_{\rm max}$ and $\Delta\nu$ scaling relations. For our work, we used $f_{\nu_{\rm max}}$ =1.0, and $f_{\Delta\nu}$ was calculated from \texttt{ASFGRID} code by \citet{2016Sharma}. The solar reference values are $\nu_{max,\odot}$ = 135.1 $\mu$Hz, $\Delta\nu_{\odot}$ =3090 $\mu$Hz and $T_{\rm eff,\odot}$ = 5777~K respectively. The asteroseismic parameters $\nu_{\rm max}$, $\Delta\nu$ and $T_{\rm eff}$ are taken from \citet{2018Yu}. We have divided the RC sample into two broad groups based on their mass: the low mass ($\leq$ 2~M$_{\odot}$) RC giants or the primary RC giants (pRCs) and high mass ($>$ 2~M$_{\odot}$) RC giants or secondary RC giants (sRCs). Similarly, we divided the RGB sample into two mass groups:  massive RGB and low-mass RGB giants. We used  demarcation at 2~M$_{\odot}$ as only giants below this mass limit are expected to develop degenerate cores on RGB. 
There are 668 pRCs, 109 sRCs, 10 massive RGBs and 453 low-mass RGBs. A significantly lesser number of high-mass giants in the sample may be due to evolutionary time scales as high-mass giants evolve much faster compared to lower-mass giants. 

\subsection{Li abundances}

We have extracted spectra of the entire sample of 1240 giants from the LAMOST DR7 survey. Most of the spectra are of good quality with S/N $\geq$ 50. Few spectra have lesser S/N but are sufficient for deriving abundances using synthesis. All spectra are continuum fitted and RV-corrected using the estimated radial velocity data from the LAMOST catalogue using tasks in \texttt{IRAF}.  We used the spectral synthesis method to account for the blending of lines with the main Li resonance line at 6707~\AA. The stellar parameters T$_{\rm eff}$, log $g$ and $[Fe/H]$ are adapted from \citet{2018Yu} catalogue. The values of microturbulent velocity ( $\xi$)
have been derived using empirical relation for giants with $[Fe/H]$ $>$ -1.0~dex \citep{2018Holtzman}

\begin{equation*}
\xi = 10^{0.226-0.0228\log g + 0.0297(\log g)^2-0.0113(\log g)^3}
\end{equation*}
and for giants with $[Fe/H]$ $\leq$ -1.0~dex \citep{2016Garcia}
\begin{equation*}
\xi = 2.478 - 0.325\log g
\end{equation*}

\begin{deluxetable*}{ccccccccccc}
\tabletypesize{\scriptsize}
\tablewidth{0.99\textwidth}
\tablecaption{ Derived and adopted parameters of sample giants \label{tab:1}}
\tablehead{\colhead{KIC}&\colhead{T$_{eff}$}&\colhead{log {\textit g}}&\colhead{[Fe/H]}&\colhead{$\xi$}&\colhead{Mass}&\colhead{A(Li)$_{LTE}$}&\colhead{$\Delta_{NLTE}$}&\colhead{$\Delta$ $\nu$}&\colhead{$\Delta$P ($\Delta \Pi_1)$}\tablenotemark{a}&\colhead{Evol.}\tablenotemark{b}\\
\colhead{}&\colhead{(K)}&\colhead{(dex)}&\colhead{(dex)}&\colhead{(km/s)}&\colhead{(M$_{\odot}$)}&\colhead{(dex)}&\colhead{(dex)}&\colhead{($\mu$Hz)}&\colhead{(s)}&\colhead{Status}}
\startdata
4136835  & 4909±80  & 2.807±0.008 & -0.62±0.15 & 1.4±0.001   & 1.47±0.08 & 0.99±0.2  & 0.07  & 7.158±0.011  & 73.67±1.49(1) & 0  \\
4137210  & 4862±80  & 2.978±0.006 & -0.35±0.15 & 1.328±0.001 & 1.35±0.07 & 1.17±0.2  & 0.15  & 9.841±0.018  & 74.3±0.65(2) & 0   \\
4243803  & 4646±100 & 2.097±0.033 & 0.05±0.15  & 1.602±0.002 & 2.39±0.55 & 0.88±0.17 & 0.32  & 1.882±0.021  & 149.75±5.17(1) & 0 \\
4345370  & 4872±100 & 2.425±0.007 & -1.17±0.15 & 1.69±0.002  & 0.96±0.05 & 1.12±0.19 & 0.11  & 4.05±0.012   & 76.08±3.03(1) & 0  \\
4346319  & 4815±100 & 2.578±0.008 & 0.2±0.15   & 1.483±0.001 & 1.97±0.14 & 1.41±0.18 & 0.15  & 4.615±0.045  & 315.1±4.53(2) & 1   \\
4346893  & 4610±100 & 2.419±0.011 & 0.39±0.15  & 1.53±0.001  & 1.23±0.11 & 1.02±0.17 & 0.32  & 3.926±0.032  & 289.4±2.86(2) & 1  \\
4446405  & 4846±100 & 2.688±0.008 & -0.13±0.15 & 1.445±0.001 & 1.58±0.09 & 1.37±0.19 & 0.15  & 5.75±0.019   & 81.04±6.15(1)  & 0 \\
4633909  & 4753±151 & 2.374±0.012 & -2.44±0.3  & 1.706±0.004 & 0.8±0.09  & 0.69±0.2  & 0.06  & 3.984±0.062  & 315.1±2.88(2) & 1  \\
4634108  & 4799±100 & 2.642±0.007 & -0.03±0.15 & 1.461±0.001 & 1.27±0.07 & 1.05±0.21 & 0.15  & 5.618±0.015  & 72.43±37.85(1)  & 0\\
4634310  & 4748±100 & 2.448±0.013 & 0.22±0.15  & 1.522±0.001 & 1.37±0.14 & 1.06±0.2  & 0.32  & 4.027±0.035  & 283.2±2.76(2) & 1  \\
\enddata
\tablenotetext{a}{(1) $\Delta$P (This work) (2) $\Delta \Pi_1$ \citep{2016Vrard}}
\tablenotetext{b}{0 - RGB, 1 - pRC, 2 sRC}
\tablecomments{Only a portion of this table is shown here to demonstrate its form and content. A machine-readable version of the full table is available.}
\end{deluxetable*}

The required line list and associated atomic and molecular data were collated by the {\emph{linemake}} code \citep{2021Placco} around the Li~I line at 6707.8~\AA.  Local thermodynamic equilibrium (LTE) model atmospheres were generated from ATLAS9 code \citep{2003Castelli} for the adopted atmospheric parameters. A series of synthetic spectra were generated 
using the updated 2019 version of the radiative transfer code MOOG \citep{1973Sneden} for each programme star by changing Li abundances. The predicted spectra were then matched with the observed spectra. Li abundance of the best-matched (least $\chi$-square) computed spectrum was taken as the star's Li abundance. S/N of spectra were adapted from the mean R-band S/N supplied by LAMOST stellar parameter pipeline (LASP) \citep{2015Xiang}. To estimate the effect of S/N on abundance measurements, we calculated errors in equivalent widths using Cayrel's formula \citep{1988Cayrel}. The weakest lines in our spectra (S/N range from 30-838) that can be detected have EWs 0.62 - 17.5 m\r{A}. We used a 3$\sigma$ limit on the EWs for reliable abundance measurements. This renders a detection limit of EW = 1.88-52.4 m\r{A} equivalent to A(Li) limits of 0.2-0.9 dex depending on T$_{eff}$ and log $g$. All sample giants have A(Li) well above the detection threshold.  A sample synthetic spectra comparison with the observed spectra is given in Fig. \ref{fig:2}. The third spectrum in the panel has one of the lowest S/N $\approx$ 30. Since these giants are cool the Li resonance line is normally very strong even for giants with moderate Li abundances. Uncertainties in the Li abundance are calculated using a quadratic sum of uncertainties attributable to the spectral quality and stellar parameters and are estimated as follows:

$$
\resizebox{0.45\textwidth}{!}
{$\Delta A(Li) = \sqrt{\Delta A(Li)_{S/N}^2 + \Delta A(Li)_{T_{eff}}^2 + \Delta A(Li)_{[Fe/H]}^2 + \Delta A(Li)_{\xi}^2}$}$$
where $$\Delta A(Li)_{S/N} = a_0 +a_1 \times S/N$$ is due to uncertainties in SNR,
$$\Delta A(Li)_{T_{eff}}=b_0+b_1\times T_{eff}+b_2\times T_{eff}^2+b_3\times T_{eff}^3$$
is due to uncertainties in T$_{\rm eff}$,
$$
\resizebox{0.45\textwidth}{!}
{$\Delta A(Li)_{[Fe/H]}=c_0+c_1\times [Fe/H]+c_2\times {[Fe/H]}^2+c_3\times {[Fe/H]}^3$}$$
is due to uncertainties in [Fe/H] and
$$\Delta A(Li)_{\xi}=d_0+d_1\times\xi$$

is due to uncertainties in $\xi$. Here $a_i,b_i,c_i,d_i$ are polynomial coefficients adapted from \citet{2021Gao}. We also provided NLTE corrections for the Li abundances utilizing $\Delta_{\rm NLTE}$ values from \citet{2009Lind}. There are 18 giants in our sample for which Li abundances were derived from high-resolution spectra in the literature \citep{2021Yan}. The mean difference between ours and those in the literature is 0.35 dex with $\sigma$ = 0.02. Evolutionary state of KIC 9907856 was undetermined by \citet{2021Yan} which we classified as a pRC from its period spacing. KIC 9596106 which was classified as RGB is also established to be a pRC star. The remaining 16 stars have the same evolutionary states as determined by \citet{2021Yan}

\section{Discussion}\label{sec:Discussion}

\begin{figure}
\includegraphics[width=\columnwidth]{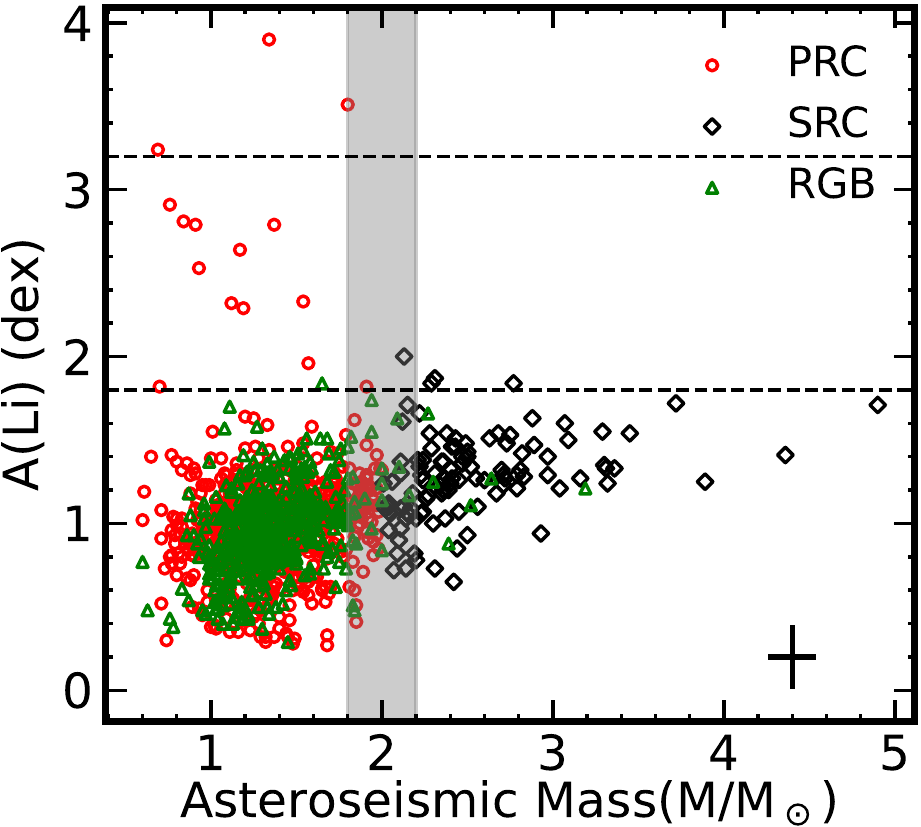}
\caption{Li abundances of pRC, sRC and RGB giants as a function of stellar mass. Horizontal lines mark the first dredge-up theoretical upper limit of A(Li) = 1.8 dex for a giant with M = 1.5~M$_{\odot}$ and A(Li) = 3.2~dex for the super Li-rich giants. The error cross at bottom right represents mean errors in A(Li) and mass}
\label{fig:3}
\end{figure}

Abundances of Li as a function of stellar mass for the entire sample are shown in Fig. \ref{fig:3}. The transition mass M$_{\rm HeF}$ = 2~M$_{\odot}$ divides the giants into two groups: a) pRCs ($\leq$ 2~M$_{\odot}$), which develop degeneracy at the core while evolving on RGB, and b) sRCs ($>$ 2~M$_{\odot}$), which develop sufficient core temperatures to burn He at the core in convective conditions while evolving on RGB, meaning no He-flash in high-mass giants. In the case of low-mass giants, at the RGB tip, ignition of He near the central core results in a thermal runaway with He-flashes generating massive energy. Note, the adopted 2~M$_{\odot}$ demarcation is an approximate average value predicted by various studies with a range of 1.8 to 2.2~M$_{\odot}$ \citep{1992Chiosi,1999Girardi}. The vertical region in Fig.~\ref{fig:3} indicates a possible range of masses that could separate pRC and sRC stars.  (See Appendix \ref{appendix:B}). Though predictions suggest He-flash generates a huge amount of energy only a part of it goes into lifting the H-burning shell upwards and hence causes a sudden drop in luminosity. The bulk of the energy goes into removing the central degeneracy. Post-He-flash, stars settle at the RC with He-burning at the core in convective conditions.

The key result from Fig.~\ref{fig:3} is that none of the sRC giants ($>$ 2~M$_{\odot}$) shows A(Li) values more than that expected from the first dredge-up.  The mean errors in A(Li) = 0.19 dex and mass = 0.14 M$_{\odot}$ are indicated by an error cross in Fig. \ref{fig:3}. We have drawn a horizontal line at A(Li) = 1.8~dex, the expected maximum first dredge-up value for a star of mass 1.5~M$_{\odot}$ \citep{1967Iben}. We found similar maximum A(Li) values for high mass giants by computing models using Modules for Experiments in Stellar Astrophysics (MESA), an open-source 1~D stellar evolution code \citep{2011Paxton,2019Paxton}, based on the study by \citet{2020Schwab}. Models are constructed for solar metallicity, as most of our giants are close to [Fe/H] = 0.0. Post first dredge-up, however, the models of high-mass giants show further depletion of Li as giants evolve to the sRC phase though at a much slower rate than their low-mass pRC counterparts which undergo extra-mixing during stars' luminosity bump evolution \citep{2021Deepak}. 
 
\begin{figure}[!ht]
\hspace*{-0.3cm}
\includegraphics[width=\columnwidth]{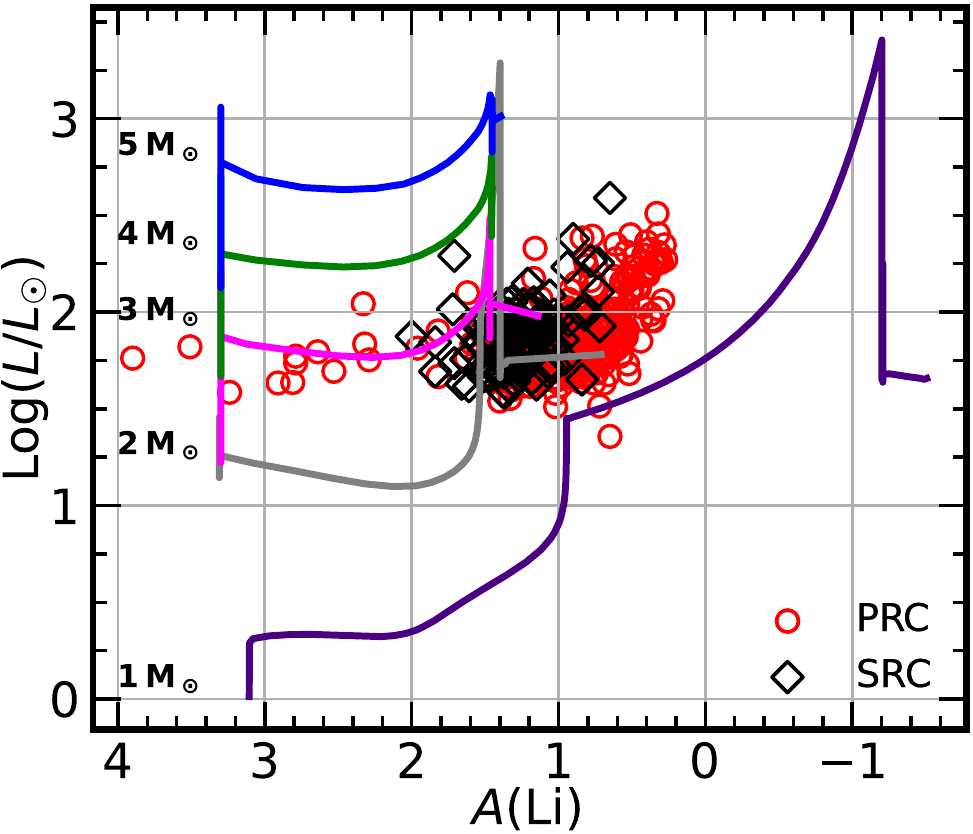}
\caption{ The pRC and sRC giants in a Luminosity and A(Li) plot superposed with MESA models of 1-5~M$_{\odot}$ giants. Note many LRGs including SLRGs among pRCs and none among sRCs. Models show pRCs suffer severe depletion of Li during 1st dredge-up and also at the luminosity bump.}
\label{fig:4}
\end{figure}

The computed models for sRC giants yield A(Li) values between $\approx$ 0.7 - 1.6 dex, depending on mass (see Fig.~\ref{fig:4}). The expected A(Li) values agree well with the observed values of sRC giants. However, three sRC giants %($\sim$ 18 \%) 
show A(Li) values slightly higher, by $\sim$ 0.1-0.3~dex than the maximum model predictions of A(Li) = 1.6~dex. Given the uncertainties in the observed values of about 0.2~dex, the slightly higher values of Li in these giants may not be attributed to fresh enhancement but to  their evolution. Also, High-mass giants begin burning He at the center well before they settle in the RC phase which means some of these giants still can have slightly higher Li abundance than those settled at RC (see Fig. 4).

Importantly, unlike pRCs, none of the 109 sRCs in our sample are super Li-rich ( A(Li) = 3.2~dex). Slightly higher A(Li) (after taking into account estimated $\approx$ 0.2~dex measurement uncertainty) for a couple of sRC giants is most probably due to insufficient mixing rather than enhancement. The observed range of A(Li) values among sRC giants is probably due to varying levels of depletion and, of course, the initial values of Li with which stars might have evolved off the main sequence. Further, one could notice from the data in Fig. \ref{fig:3} and the models in Fig. \ref{fig:4} that the range of A(Li) among sRC giants is relatively smaller compared to pRC giants. It is known that low-mass giants undergo severe depletion of Li due to  1st dredge-up and extra-mixing near the luminosity bump. The high mass giants neither have sufficient time for Li depletion, as they evolve faster nor do they have extra-mixing at luminosity bump. As shown in Fig. \ref{fig:4}, the 1~M$_{\odot}$ model suffers significant Li depletion both during 1st dredge-up and luminosity bump evolution. The lack of LRGs among sRCs provides another clue that He-flash is the key source of Li-enrichment among pRCs. 

%The present study aims to understand whether sRC giants show evidence of Li enhancement. 

Previous attempts by \citet{2019Deepak, 2021Deepak} to understand Li distribution as a function of mass could not resolve the issue either due to small sample size or unreliable mass determinations using stellar evolutionary tracks. Most mass tracks overlap in RC region in HRD  making it challenging to derive masses solely based on evolutionary tracks. Our study overcame these shortcomings by assembling a large sample common among Kepler and LAMOST catalogues. The studies by \citet{2021Martell, 2022Zhou}, however, suggest  LRGs are a diverse population found among sRCs, pRCs and also among  RGB giants. The main differences between theirs and our study are; the primary sample source and the method used to classify giants as RCs and RGBs. Our sample is sourced from \citet{2018Yu} catalogue of Kepler giants with oscillations identified. All giants in our study have direct measurement of $\nu_{\rm max}$, a key asteroseismic parameter for evolutionary phase determination and also have $\rm \Delta P$ values measured directly from asteroseismic data. On the other hand, the primary source of the sample for the \cite{2021Martell} study is GALAH, and hence most of their sample giants do not have asteroseismic data. They identified giants as RCs and RGBs using stellar parameters which are proven to be ineffective for obvious reasons as RC and upper RGB regions overlap in the HRD. The study by \cite{2022Zhou} also shows a few SLRs among sRCs. Their sample is primarily drawn from the LAMOST survey similar to ours but  the classification of RCs and RGBs is based on neural networks. Unfortunately, there are no common giants between \cite{2021Martell} and ours. \cite{2022Zhou} didn't publish their sample set.
The other difference between ours and \citet{2021Martell} is their lower mass cut-off of 1.7~M${\odot}$. However, this does not make difference to the results shown in Fig. \ref{fig:3}. Among the many LRGs, only two are close to the adopted M$_{HeF}$ = 2~M$_{\odot}$, an upper limit for the He-flash phenomenon which is possibly due to metallicity effect (see Appendix~\ref{appendix:B}). It may be possible that giants might have lost 0.2~M$_{\odot}$ to 0.3~M$_{\odot}$ as they evolved to the RC phase (see \citet{2022Chaname}). This suggests that stars with an initial mass of about $\leq$ 2.2~M$_{\odot}$ undergo He-flash implying only RC giants with a current mass of about 2~M$_{\odot}$ or less show Li enrichment.

\par
%In this work, we have synthesised 1228 stars both along the RGB and RC phases to show how Li varies according to their masses with particular attention on the low mass primary RC and higher mass ( $>$ 2$M_{\odot}$ ) secondary RC stars. In Fig. \ref{fig:3a} distribution of Li with stellar mass is shown. 
%None of the sRC stars are SLR. The highest Li in sRC stars is 1.9 $\pm$ 0.18 dex (NLTE correction - 0.13 dex). It is interesting to note that while $\sim$ 64.3\% pRC stars (374/582) are Li-rich i.e. A(Li) $>$ 1 dex \citep{2019Singh}, a whopping $\sim$ 91.7\% sRC stars (88/96) are Li-rich. It is worth noting that average A(Li) in Li-rich pRC (1.29 $\pm$ 0.19 dex) and sRC stars (1.35 $\pm$ 0.17 dex) are in close range.
%\subsection{MESA modelling for sRC stars}
%We used stellar evolution models to illustrate lower Li abundances in sRC stars. A number of stellar models were estimated spanning the mass ranges covered by our sRC sample. A solar mass model was also considered to reflect the disparity in Li evolution in stars with different mass ranges. Since our pRC and sRC stars peak at metallicities of [Fe/H]= 0 \& 0.1Z$_{\odot}$ respectively, only solar metallicity was investigated. Models were constructed using release r15140 of Modules for Experiments in Stellar Astrophysics (\texttt{MESA}), an open-source 1D stellar evolution code \citep{2011Paxton,2019Paxton}. The source codes compiled by \citet{2020Schwab} were used with some modifications as demonstrated in \citet{2021Deepak}. 
\par

\section{Conclusion}
We have used a large sample of RGB and RC stars with evolutionary phases classified using the asteroseismic diagram of $\rm \Delta P$ - $\Delta \nu$ based on direct measurements of Kepler light curves. For the entire sample of 1240 giants (777 RCs and 463 RGBs), we derived Li abundances using spectral synthesis. We found no evidence of Li-enrichment among sRCs. The observations conform with the theoretical A(Li) predictions for sRC giants of M $>$ 2~M$_{\odot}$.
We found all pRC giants, including  three SLRGs and 11 giants with A(Li) $>$ 1.8~dex, whose A(Li) values are  much higher than that expected from models and their counterparts on upper RGB.  Also, we found no giant on RGB with Li abundance more than the upper limit of A(Li) = 1.8~dex expected from models for low-mass giants within the uncertainties.
The lack of  Li-rich giants among sRC stars is another clue that the He-flash, which only occurs in low-mass giants, is the potential site for Li-enrichment among low-mass giants. This result further strengthens the growing evidence that Li-enrichment occurs during He-flash. However, the transport process and mixing mechanism are yet to be explored. It would be worthwhile to combine carbon isotopic ratios ($\mathstrut^{12}C$/$\mathstrut^{13}C$) with lithium studies to understand mixing mechanisms. 
%Also, a large-scale survey of giants classified as RCs and RGBs based on direct measurements from asteroseismology using Kepler and TESS data will help to determine whether Li enhancement in giants has single origin or multiple origins as some claim. 
\\

%pRC stars experience severe Li depletion post luminosity-bump due to extra mixing mechanisms. The bump is absent in higher mass stars. Hence Lithium depletion is much slower which explains the narrow range of A(Li) in sRC stars. 
 %It has been established that in pRCs, rotational and thermohaline mixing both contribute in changing Li abundance\citep{2010Charbonnel}.
%Our study should be complemented with rotational velocity measurements and its effects on the secondary red clump.  \par

%We thank anonymous referees for their constructive comments.
This work has used data from the \citet{2018Yu} catalogue available at \url{http://vizier.u-strasbg.fr/viz-bin/VizieR?-source=J/ApJS/236/42}. All spectra were taken from LAMOST public data release 7, operated and managed by National Astronomical Observatories, Chinese Academy of Sciences. We are grateful to the team of Stellar Classification Program (SCP) for the Kepler Mission. We acknowledge utilization of \texttt{MESA} models assembled by Josiah Schwab. Schwab's input and output files are publicly available on Zenodo at doi: \dataset[10.5281/zenodo.3960434]{\doi{10.5281/zenodo.3960434}}.\par
\software{\texttt{ASFGRID} \citep{2016Sharma}, \texttt{IRAF} (\url{https://ui.adsabs.harvard.edu/abs/1993ASPC...52..173T}), \texttt{MOOG} \citep{1973Sneden}, \texttt{MESA} \citep{2011Paxton,2019Paxton}, \texttt{MESASDK} 20.3.1 \citep{richard_townsend_2020_3706650}, \texttt{py\_mesa\_reader} \citep{bill_wolf_2017_826958}}

\begin{appendix}
\begin{figure*}[!ht]
\centering
\includegraphics[width=\textwidth]{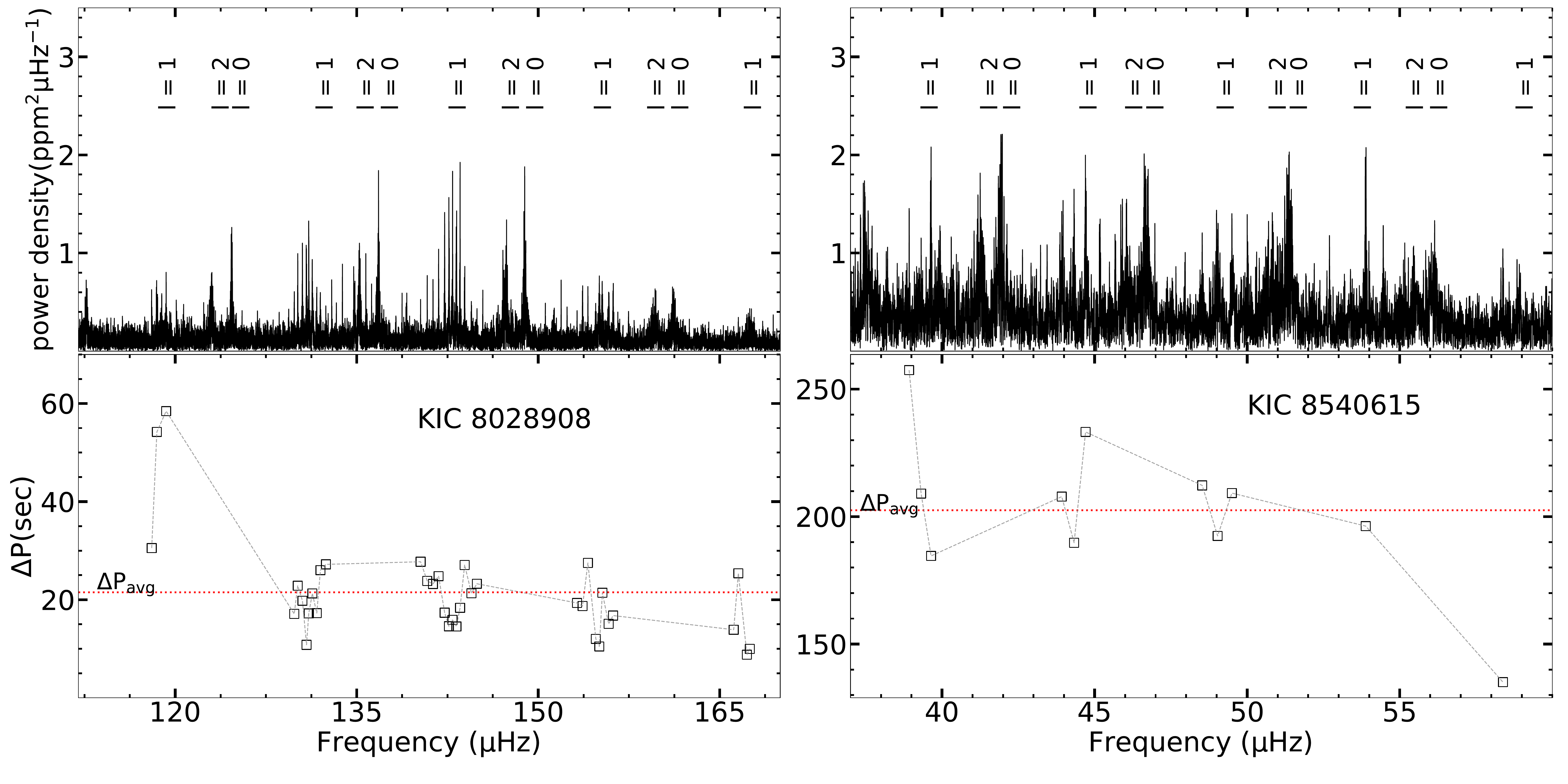}
\caption{Measurement of period spacing. The top left panel is the PDS of RGB star KIC~8028908 and the right panel is the PDS of RC star KIC~8540615. Oscillation modes of l = 0, 1, 2 are identified and marked in the figure. The bottom panel is a demonstration of the measurement of mixed mode (l = 1) period spacing ($\rm \Delta P$). The red dotted horizontal line is the average value of period spacing in each bottom panel.}
\label{fig:period-spacing}
\end{figure*}
\section{Measurement of mixed-mode period spacing}\label{appendix:A}
Although all of our stars have known evolutionary phases in \cite{2018Yu}, we derived period spacing to precisely infer the evolutionary phase based on the location of a star in the asteroseismic plot of $\rm \Delta P$ - $\Delta \nu$ \citep{Bedding2011}. Evolved stars show a rich spectrum of oscillation modes in the power density spectra (PDS) which are radial, dipole and quadruple modes. Dipole modes of evolved stars have mixed natures i.e. they arise from coupling between $p$ - modes in the envelope and $g$ - modes in the core. Consecutive radial modes are equally spaced in frequency and mixed dipole modes are approximately equally spaced in period \citep{Tassoul1980}. Spacing of the period between consecutive mixed dipole modes has been used to identify different evolutionary phases of stars \citep{Bedding2011, Mosser2012, Stello2013}. For measurement of period spacing, we retrieved Kepler photometric data from MAST archive \footnote{\url{https://mast.stsci.edu/portal/Mashup/Clients/Mast/Portal.html}} using Lightkurve code \footnote{\url{https://github.com/lightkurve/lightkurve}} and converted lightcurve into frequency space following Lomb-Scargle periodogram method. We did a visual inspection of PDS for the identification of oscillation modes. Stars shows radial modes (l = 0), dipole modes (l = 1) and quadruple modes (l = 2). In each star, we identified three to five groups of mixed modes (see top panel of Figure~\ref{fig:period-spacing}) and derived period spacing from consecutive mixed dipole modes. The mean value of period spacing is adopted as period spacing of star and standard deviation as error of period spacing \citep{Bedding2011, Stello2013, 2019Singh}. 
\par
\begin{figure*}
    \centering
    \includegraphics[width=9cm]{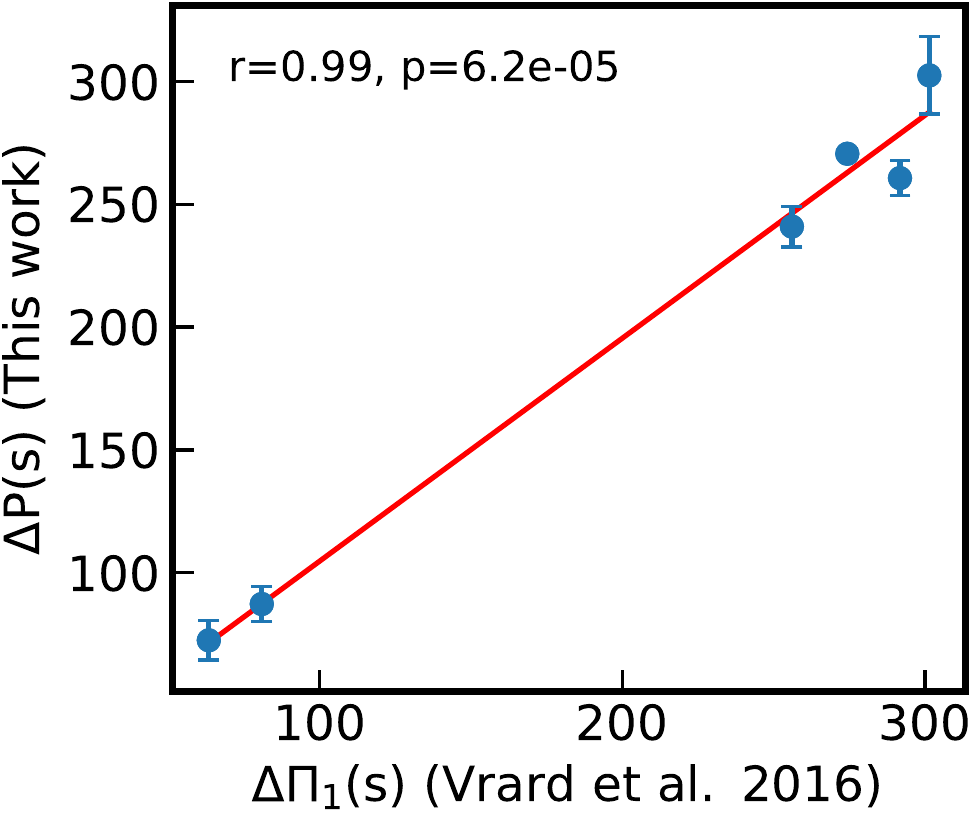}
    \caption{Average period spacing derived by us compared to values obtained by \citet{2016Vrard}}
    \label{fig:6}
\end{figure*}

To check the accuracy of our method we compared $\Delta$P values of six giants ( 2 each from RGB, pRC and sRC phase ) measured by us with those from \citet{2016Vrard} sample. As illustrated in Fig.\ref{fig:6}, our values are in good agreement as indicated by linear regression coefficients.

\section{VARIATION OF TRANSITION MASS IN FLASHING STARS}\label{appendix:B}
\begin{figure*}[!ht]
\centering
\includegraphics[width=9cm]{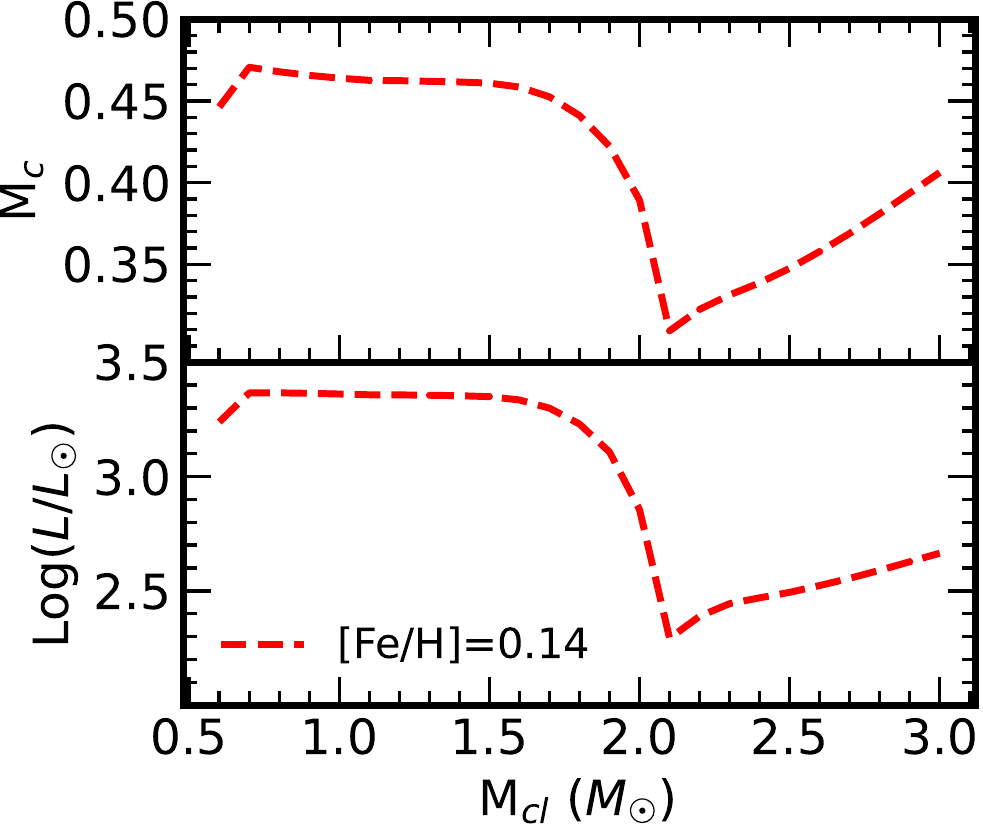}
\caption{He-core mass and Luminosity as a function of stellar mass for two metallicities for stars evolving from MS to CHeB phases.
\label{fig:5}}
\end{figure*}

The range of masses for Helium flash to occur in a star varies slightly with its composition. Masses range from 1.8 - 2.2~M$_{\odot}$ \citep{1992Chiosi} or 2-2.5~M$_{\odot}$\citep{1999Girardi}. We used a median value of 2~M$_{\odot}$ to differentiate pRC and sRC stars. In Fig.~\ref{fig:3} there is one SLR star close to M$_{\rm HeF}$ = 2~M$_{\odot}$. To confirm its evolutionary status, we have plotted Helium core mass (M$\rm _c$) and luminosity L from the main sequence to the end of the core Helium burning phase using a MESA stellar model of [Fe/H] = 0.14. The minima is used to constrain M$_{\rm HeF}$ value. From Fig. \ref{fig:5},

\begin{itemize}
\item KIC 8879518 - M$_{Cl}$ = 1.80~M$_{\odot}$, [Fe/H] = 0.14, M~$_{\rm HeF}$ = 2.1~M$_{\odot}$
\end{itemize}
Its M$_{Cl}$ $<$ M$_{\rm HeF}$ and is confirmed to be a SLR pRC star.
\end{appendix}

\bibliography{anohita}{}
\bibliographystyle{aasjournal}

\end{document}